\documentclass[
10pt,
showpacs,preprintnumbers,nofootinbib,
amsmath,amssymb,
aps,
prl,twocolumn,groupedaddress,superscriptaddress,
showkeys
]{revtex4-1}
\pdfoutput=1
\usepackage{graphicx}
\usepackage{dcolumn}
\usepackage{bm}
\usepackage[colorlinks=true,urlcolor=blue,citecolor=blue]{hyperref}
\usepackage{color,lineno}

\begin{document}

\title{$\beta^{2}$ corrections to spherical EDF calculations for root-mean-square charge radii}

\author{B. Alex Brown and Kei Minamisono }
   \affiliation{
Department of Physics and Astronomy, FRIB Laboratory,
   Michigan State University, East Lansing, MI 42284-1321, USA}

\date{\today}

\begin{abstract}
Root-mean-square charge radii are discussed in terms of
spherical Energy Density Functional (EDF) models corrected for quadrupole
deformations. Comparisons between experiment and theory are made for
the absolute radii of all even-even nuclei, for the isotonic shift
between cadmium and tin isotopes, the isotopic shifts of the calcium
isotopes and the isotonic shift for nuclei with $  N=28  $.
We conclude that the data are well described in this approach,
except for the sharp rise just after the neutron magic numbers.

\end{abstract}

\maketitle

\section{Introduction}

Root-mean-square (rms) charge radii of nuclei provide one of
the most precise insights into nuclear structure.
New experiments are being carried out on long chains
of isotopes, and theoretical models are being improved,
see \cite{ko22}, \cite{pe21}, \cite{re17} and references therein.
Fig. (1) shows the measured rms charge radii for
even-even nuclei from calcium to tellurium.
One observes kinks in the isotopic trends
at neutron numbers $  N=28  $ and $  N=50  $, as well as kinks in the isotonic
trends at the proton numbers $  Z=28  $ and $  Z=50  $.
In particular, the isotonic kink at $  Z=50  $
results in a close spacing between the
rms charge radii of cadmium and tin isotopes
which is one of the topics of this paper.

We start with a comparison of the data to
spherical energy-density functional (EDF) calculations using the 12 Skyrme functionals from Table I 
of
\cite{br14}.
As shown in Fig. 4 of \cite{ro15}, the main difference in the
results for the rms charge radii for
these 12 functionals is correlated with the effective mass.
One group has $  m^{*}/m \approx 1.0  $. The results for $  s8  $ parameter set
which is representative of this group are shown in Fig. (2).
The other  group has $  m^{*}/m \approx 0.7-0.8  $. The results for $  s17  $ parameter set
which is representative of this group are shown in Fig. (3).
These are carried out in a spherical basis with
proton and neutron occupation numbers obtained from a
self-consistent exact-pairing calculation \cite{vo01} based on the
$  (J,T)=(0,1)  $ two-body matrix elements from
configuration-interaction Hamiltonians used for the various mass regions.
The calculations include the proton and neutron finite sizes
and the relativistic spin-orbit corrections as described in
\cite{br84}.

The difference between experiment and the spherical
calculations shows the well-known effect of nuclear deformation
in between the magic numbers of 28, 50, 82 and 126.
To account for deformation one can carry out deformed EDF calculations.
Many such calculations give results at the energy minimum for
a given $\beta$ deformation. More advanced calculations include
fluctuations around the minimum \cite{de10}.

\begin{figure}
\includegraphics[scale=0.45]{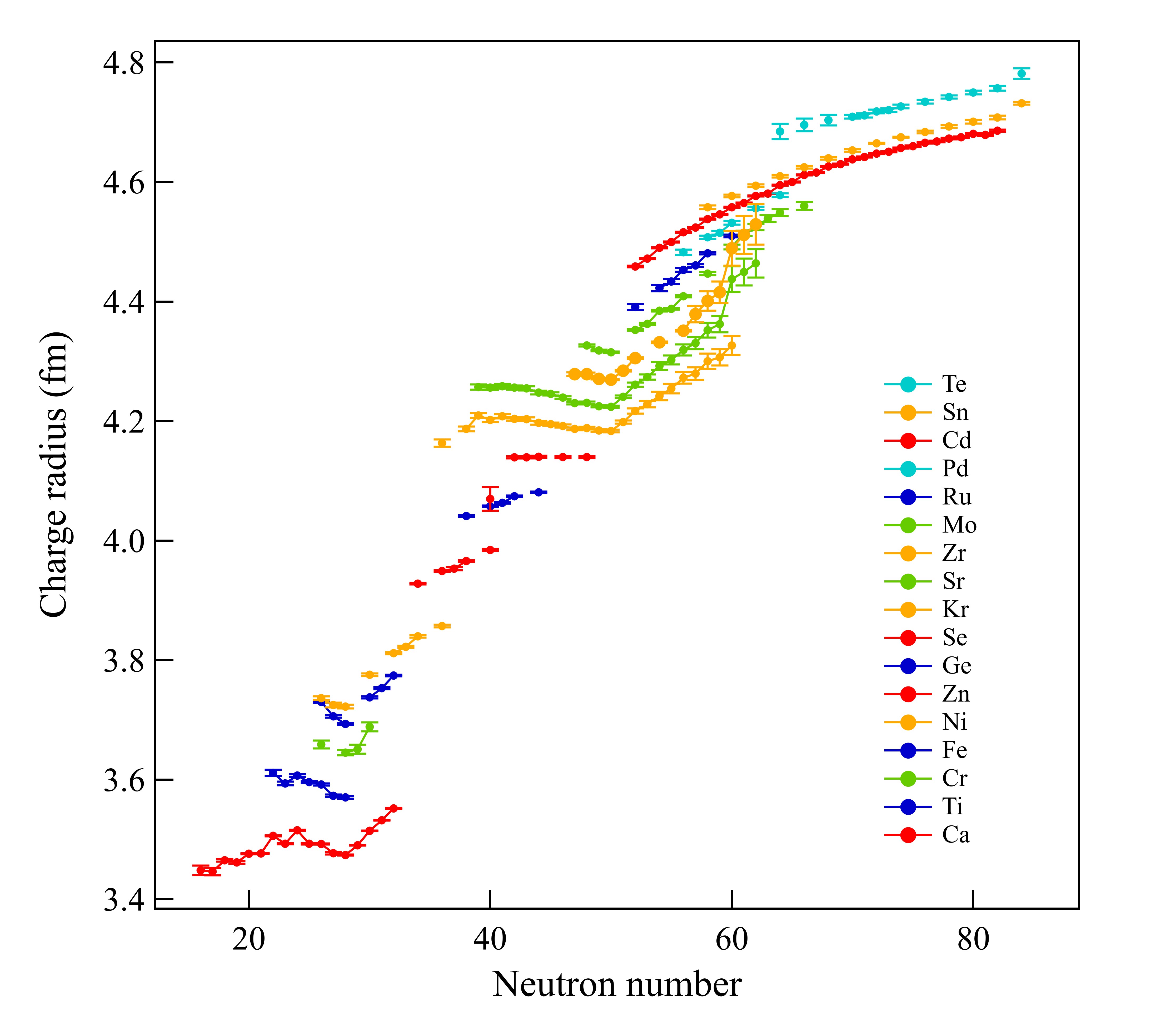}
\caption{Measured rms charge radii for even-even
nuclei from  $  Z=20  $ (calcium) to $  Z=52  $ (tellurium).
The data are taken from compilations \cite{an13} and \cite{fr04} with updated
results for Ca \cite{ga16}, Fe \cite{mi16}, Ni \cite{so22}, Cd \cite{cd18}.
and Sn \cite{sn19}.
}
\label{ (1)] }
\end{figure}

A different approach is
to use the information contained in the
reduced transition probability $  B(E\lambda )  $
for the excitation of low-lying states
to deduce an effective defomation parameter $\beta$ (this includes the
fluctuations) and then use the Bohr collective model
to evaluate the change in the rms radius taking into
account volume conservation as a function of deformation \cite{bo69}.
To order $\beta^{2}$ the mean-square radius is
$$
<r^{2}>
=\, <r^{2}>_{0} \Big[1 + \frac{5 \beta ^{2}}{4\pi } \Big],       \eqno({1})
$$
where
$$
<r^{2}>_{0} = \frac{3R_{0}^{2}}{5}       \eqno({2})
$$
is the mean-square radius with no deformation,
and
$$
\beta ^{2} =  \displaystyle\sum _{\lambda  \geq 2}  \beta _{\lambda }^{2}.       \eqno({3})
$$
The proton $\beta^{2}$ is related to the $  B(E\lambda ,\uparrow)  $ for 0$^{ + }$ to $\lambda^{ + 
}$
transitions (in units of e$^{2}$) by
$$
\beta _{\lambda }^{2} = \frac{B(E\lambda ,\uparrow)} { [\frac{3}{4\pi } Z \, R_{0}^{\lambda }]^{2} 
}.       \eqno({4})
$$
The operator for $  E2  $ is
$$
     \displaystyle\sum _{i} r_{i}^{2} Y^{(2)}(\hat{r}_{i}) e_{iq} {\rm e}       \eqno({5})
$$
where $  e_{q}  $ is the effective charge for protons ($  q=p  $)
or neutrons ($  q=n  $).
Fluctuations of the nuclear surface give corrections to the
rms charge radii that have the same form as
those of the Bohr deformed model \cite{es83}.

We will assume equal (isoscalar) proton and neutron
deformations. A more general expression for Eq. (1)
with unequal proton and neutron deformations
is given in \cite{pi21}. Results for higher-order terms in Eq. (1)
are given \cite{do84}, \cite{ah88}, \cite{be19}. For $\lambda$=2,
the next most important term is proportional
to $\beta_{2}^{3}$.
For $\mid\beta\mid$= 0.6 this term gives about a 10\% change to Eq. (1).
This term depends on the sign of $\beta$
which is often not measured, and it will not be included here.

\begin{figure}
\includegraphics[scale=0.6]{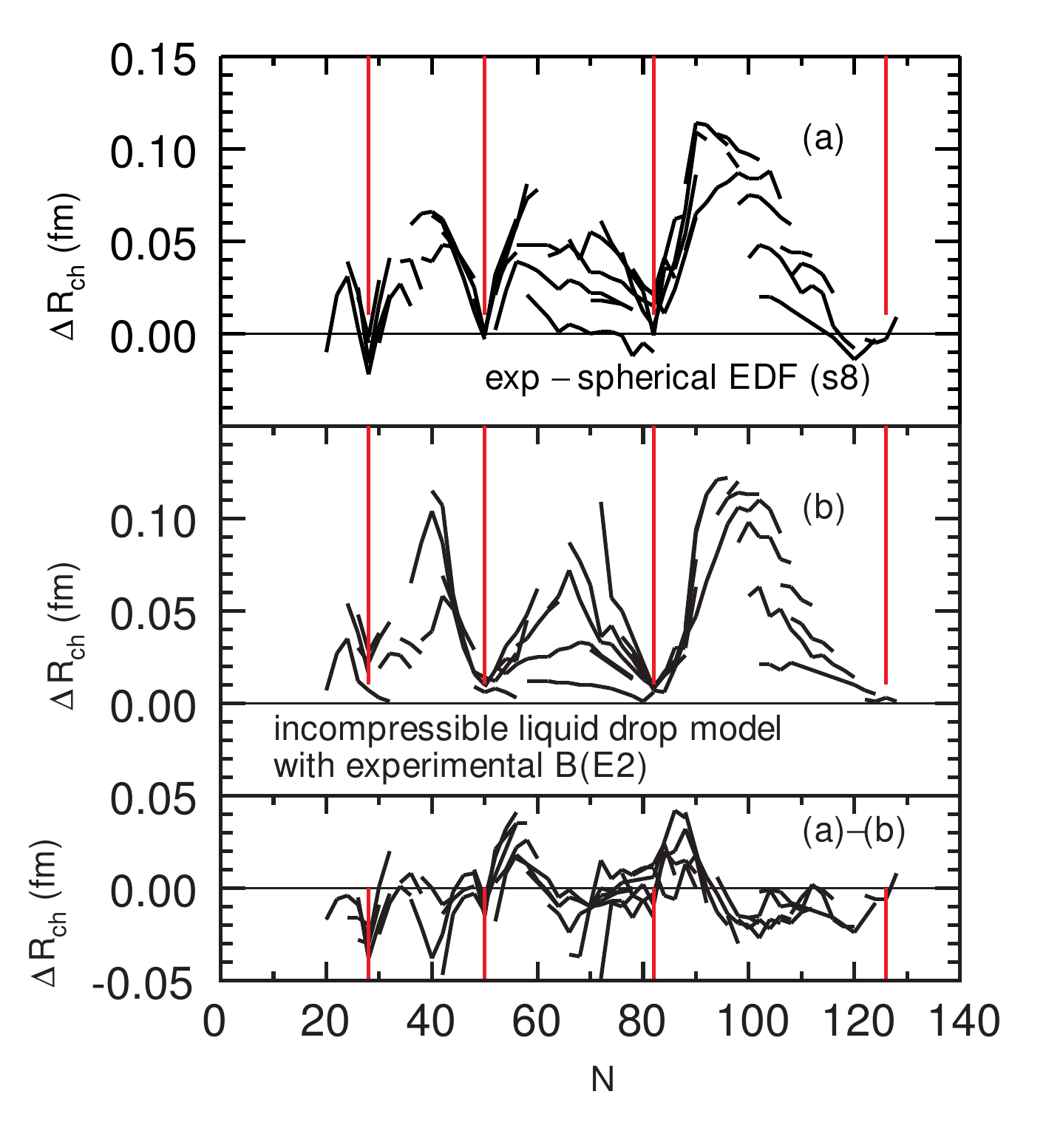}
\caption{(a) Difference between the experimental rms charge
radii and those obtained in a spherical EDF
calculation  with the $  S8  $ Skyrme functional.
(b) The increase in the rms charge radii calculated with Eq. (1)
using experimental $  B(E2)  $ values. (c)
The difference between (a) and (b).
}
\label{ (2)] }
\end{figure}

\begin{figure}
\includegraphics[scale=0.6]{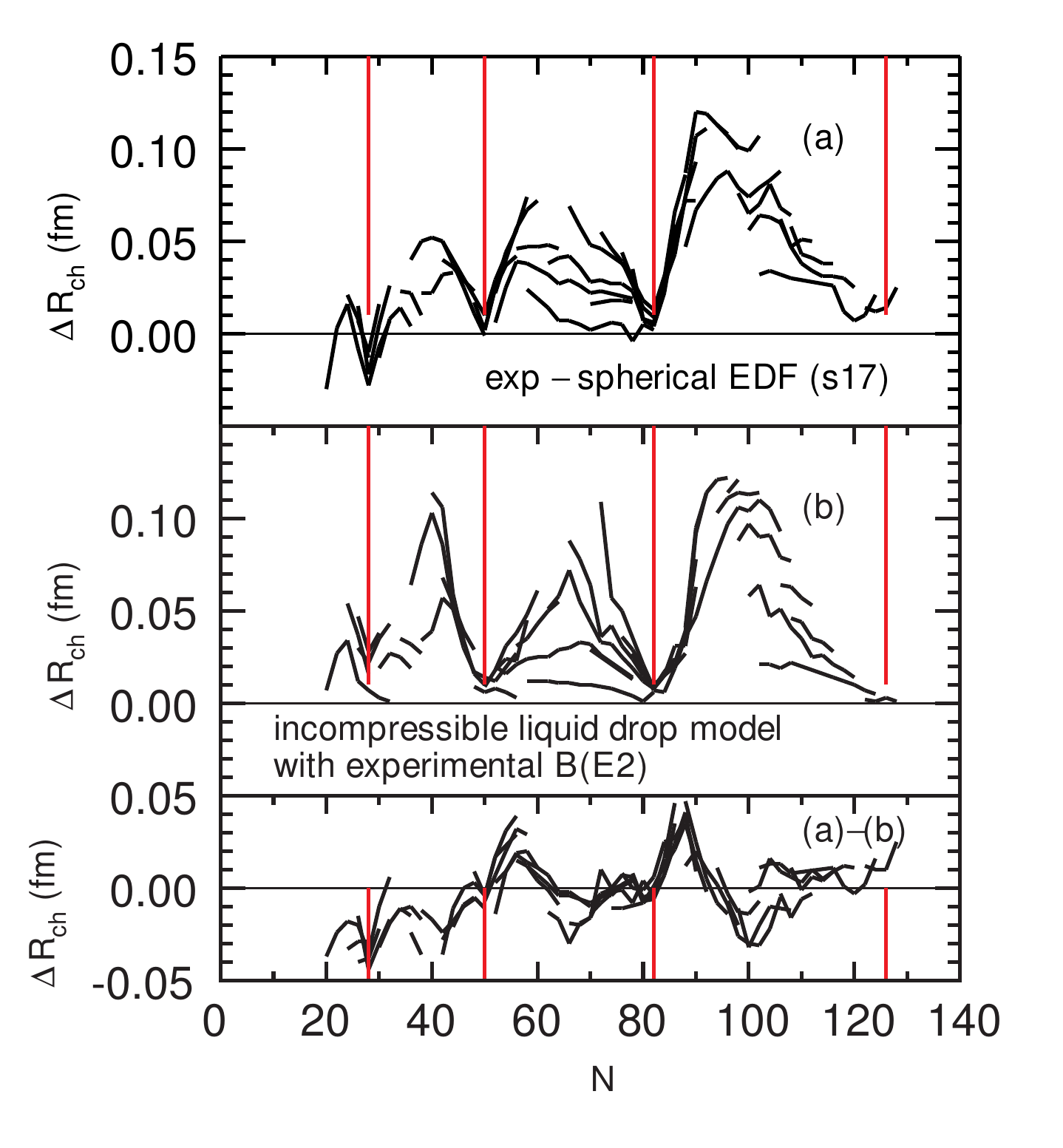}
\caption{(a) Difference between the experimental rms charge
radii and those obtained in a spherical EDF
calculation  with the $  s17  $ Skyrme functional.
(b) The increase in the rms charge radii calculated with Eq. (1)
using experimental $  B(E2)  $ values. (c)
The difference between (a) and (b).
}
\label{ (3)] }
\end{figure}

The strength functions for $  B(E2,\uparrow)  $ are dominated
by the low-lying excited 2$^{ + }$ state and the high-lying
giant quadrupole state. Low-lying excited states fall into
three groups, (a) states in well deformed nuclei,
(b) states in semi-magic nuclei formed from valence configurations
below 2-3 MeV, and (c) states in doubly-magic
nuclei formed by particle-hole excitations.

In this paper we compare experimental
rms charge radii for nuclei on or near the semi-magic numbers
with the results from three types of EDF calculations. The first is for the
spherical EDF with parameters determined from bulk properties
of doubly-magic nuclei \cite{skx}, \cite{br13}, \cite{br14},
or from a large sample of nuclear bulk properties
together with other observables resulting in the Sv-min set of
EDF paramters \cite{kl09}.
The second is the same spherical calculations but supplemented
with $\beta_{2}^{2}$ determined from experimental and/or calculated $  B(E2)  $
values of type (a) and (b) above.
The third is the Fayans-type EDF \cite{fa00}, \cite{sa11} with pairing terms
added to provide an improved description of rms charge radii
resulting in the Fy($\Delta$r) set of parameters \cite{re17}.

In the following we first consider the global comparison
for even-even nuclei using experimental data
for the $  B(E2)  $ values. We will call this
the $\beta_{2}^{2}$ correction. We also look at the relative size of
contributions to the rms radii
coming from $\lambda$=2, 3 and 4. Next we look at the specific
application to the isotonic shift betwen cadmium and tin
to compare various approaches. Then we consider the
isotopic shifts for $  Z=20  $ (calcium) and the isotonic
shifts for $  N=28  $ using experimental $  B(E2)  $
values. We also use configuration-mixing calculations
for $  Z=20  $ and $  N=28  $ to calculate  $  <Q \cdot Q>  $
for both even-even and odd-even isotopes in order
to explore the odd-even oscillations in the
rms charge radii.
At the end we discuss the problem for the increase
in charge radii after the magic numbers and
suggest possible solutions.

\section{Global Comparisons}

Using the compiled experimental $  B(E2,0^{+} \rightarrow 2^{+}_{1})  $ values
\cite{be2} the correction to the rms radii implied by Eq. (1) are
shown in Figs. (2) and (3).
The bottom panel of these figures show difference
$  R_{ch}  $(experiment) - $  R_{ch}  $(theory) - $  \Delta R  _{ch}$($\beta_{2}$).
$  \Delta R  _{ch}$($\beta_{2}$) is the  calculated correction based on $\beta_{2}^{2}$.
The increase away from the magic numbers
accounts for part of the kinks in the experimental
radii at the magic numbers. There are some remaining deviations just after the
magic numbers. For the region of the mass number $  A=100  $,
the $  B(E2)  $ for higher 2$^{ + }$ states
are typically less than about 5\% of those for the
first 2$^{ + }$ state, see for example \cite{li76}.
The contribution for the higher 2$^{ + }$ states for the calcium isotopes
and the $  Z=28  $ isotones will be discussed below.

\begin{figure}
\includegraphics[scale=0.6]{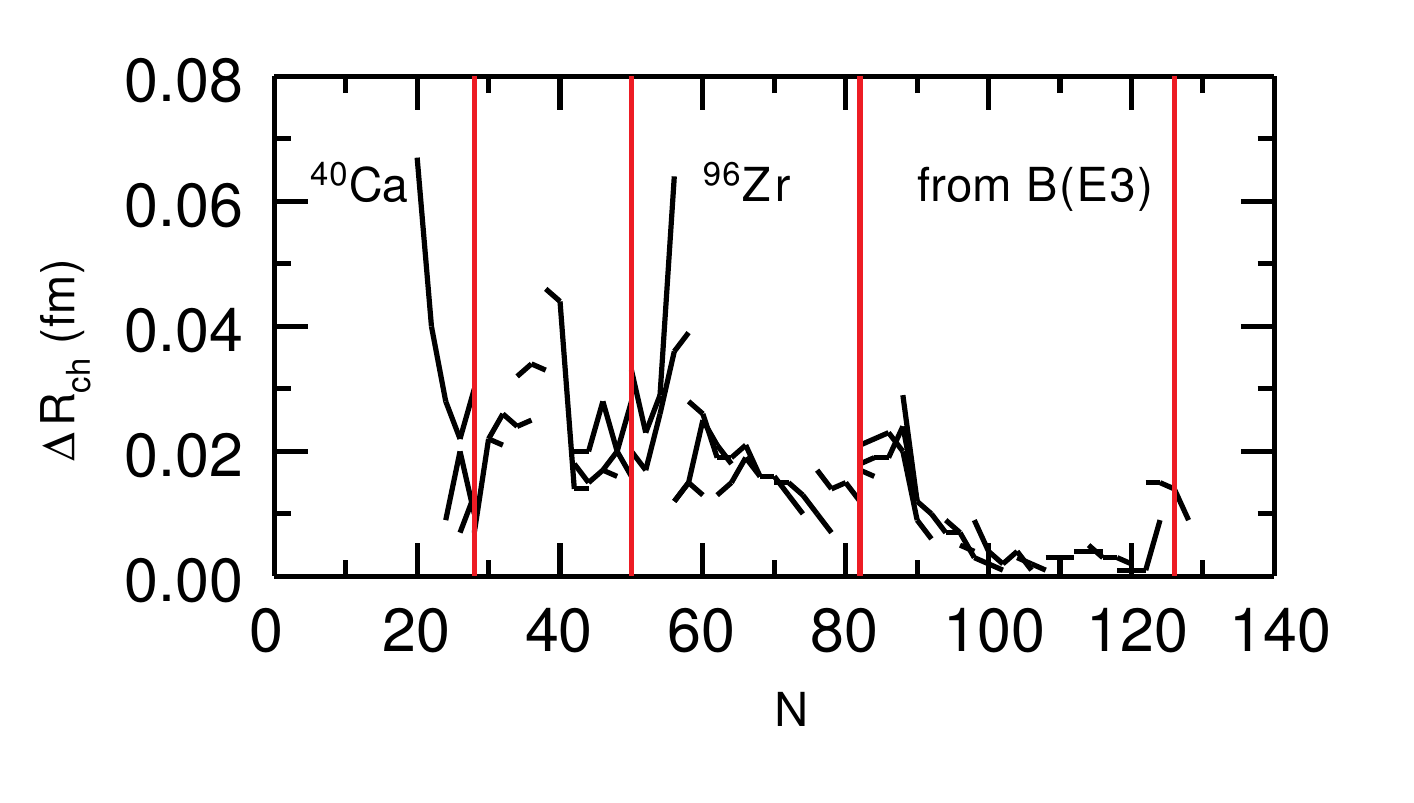}
\caption{ Increase
in the rms charge radii calculated with Eq. (1) using
experimental $  B(E3)  $ \cite{be3}.
}
\label{ (4)] }
\end{figure}

The next most important multiple is $\lambda$=3.
The increase in the rms charge radii obtained from the
experimental $  B(E3,0^{+} \rightarrow 3^{-}_{1})  $ values \cite{be3}  are shown in Fig. (4).
In contrast to the dramatic shell effects observed for the
$  B(E2)  $ in Figs. (2) and (3), the radius increase from
the $  B(E3)  $ are rather flat. For the doubly-magic nuclei
the $  B(E3)  $ are particle-hole excitation of the type (c) above.
To the extent that the EDF parameters are determined to reproduce
the rms radii of these doubly-magic nuclei, the effects of
the $  B(E3)  $ contributions are contained in the EDF calculations.
For nuclei in between the magic numbers, the $  B(E3)  $ is typically
split over many low-lying states and their total contribution to the rms
radii may be larger than that shown in Fig. (4).
The largest $  \Delta R  _{ch}$ are observed for $^{40}$Ca and $^{96}$Zr.
The implications of the increase in $  \Delta R  _{ch}$(E3) for $^{96}$Zr and
nuclei just after the
magic numbers will be discussed at the end.

Results for the radii using the $\beta_{2}$ and $\beta_{4}$ obtained from the BSk27 EDF calculations
\cite{go13} are shown in Fig. (5).
$\beta_{4}$ accounts for about a third of the difference just after $  N=82  $
shown in the bottom panels of Fig. (2) and (3).

\section{Isotonic Shifts for the Cadmium and Tin Isotopes}

Modifications have been made to the Skyrme functionals
to account for odd-even oscillators in the rms charge radii
and the increase in radii after the
magic numbers. The Fayans functionals add a pairing-type
term to spherical EDF \cite{fa00}, \cite{sa11}. The M3Y-P6a functional adds
density-dependent spin-orbit and pairing terms to the
M3Y-type functional \cite{na15}.

As an example for isotonic shift in the rms charge radii,
we consider the results for the cadmium and tin isotopes.
These are shown in Fig. (6) using the experimental
data and calculations given in \cite{cd18} for cadmium and \cite{sn19} for tin.
The experimental data are compared to the spherical calculations with the
Sv-min \cite{kl09} EDF calculations given in the experimental papers \cite{cd18}, \cite{sn19}.
All 12 Skyrme functionals given in \cite{br14} give isotonic shifts
similar to those shown for Sv-min. The experimental isotonic shift
is closer to zero than to the standard Skyrme results.

In Fig. (6) The isotonic shifts are also compared to the results from the
Fy($\Delta$r,HFB) functional given in \cite{cd18}, \cite{sn19},
and to the results obtained by adding the
$\beta_{2}^{2}$ corrections
using the experimental $  B(E2)  $ values from \cite{pr17}
to the Sv-min values.
The error bars
shown in Fig. (6) are from those given in \cite{pr17} for the
$  B(E2)  $ and from those given in \cite{cd18} and \cite{sn19}
for the rms radii.
The  Sv-min plus $\beta_{2}^{2}$ corrections provide a reasonbaly good description of
the data.
The remaining differences between theory
and experiment may be due to: (a) systematic uncertainties  in the
charge radii, (b) systematic uncertainties  in the $  B(E2)  $ extracted from Coulex
and/or lifetime measurements,
(c) deficiencies in the Sv-min EDF, or (d) $\beta_{4}^{2}$ corrections.

\section{Isotopic Shifts for the Calcium Isotopes}

One of the most famous and challenging data for rms charge
radii is that for the calcium isotopes where there is a strong
odd-even oscillation in the rms charge radii with
$^{42,44,46}$Ca being relatively large compared to those for
the "closed-shell" nuclei $^{40}$Ca and $^{48}$Ca.
It is notable that the experimental rms charge radii
of $^{40}$Ca and $^{48}$Ca are nearly the same \cite{ga16}.
This data has led to many theoretical ideas \cite{ta84},
\cite{za10}, \cite{ho94}, \cite{za87}, \cite{ca01}, \cite{za85}, \cite{ba85},
\cite{fa00}, \cite{sa11}.
It was noted
by Talmi \cite{ta84} that a two-body effective operator for a correction to
rms radii contains odd-even oscillations.
The increase in the rms charge radii is correlated
with an increase in the matter rms radii \cite{gi92}.
The increase in the rms radii for $^{42,44,46}$Ca
relative to $^{40}$Ca and $^{48}$Ca
are in relatively good agreement with $\beta_{2}^{2}$ corrections
obtained from the experimental $  B(E2)  $ values as shown in
panels (a) and (b) of Fig. (7). The data are
for $^{42}$Ca \cite{ca42}, $^{44}$Ca \cite{ca44}, $^{46}$Ca
and $^{50}$Ca \cite{ca50}. For $^{42}$Ca we include the the $  B(E2)  $
for to the 2$^{ + }_{2}$ state for which
$  B(E2,0^{+}_{1} \rightarrow 2^{+}_{2})/B(E2,0^{+}_{1} \rightarrow 2^{+}_{1}) \approx 0.1  $
\cite{ca42}.
The theoretical results from the
Fy($\Delta$r,HFB) calculations given in \cite{mi19} are
shown in panel (c) of Fig. (7).

\begin{figure}
\includegraphics[scale=0.6]{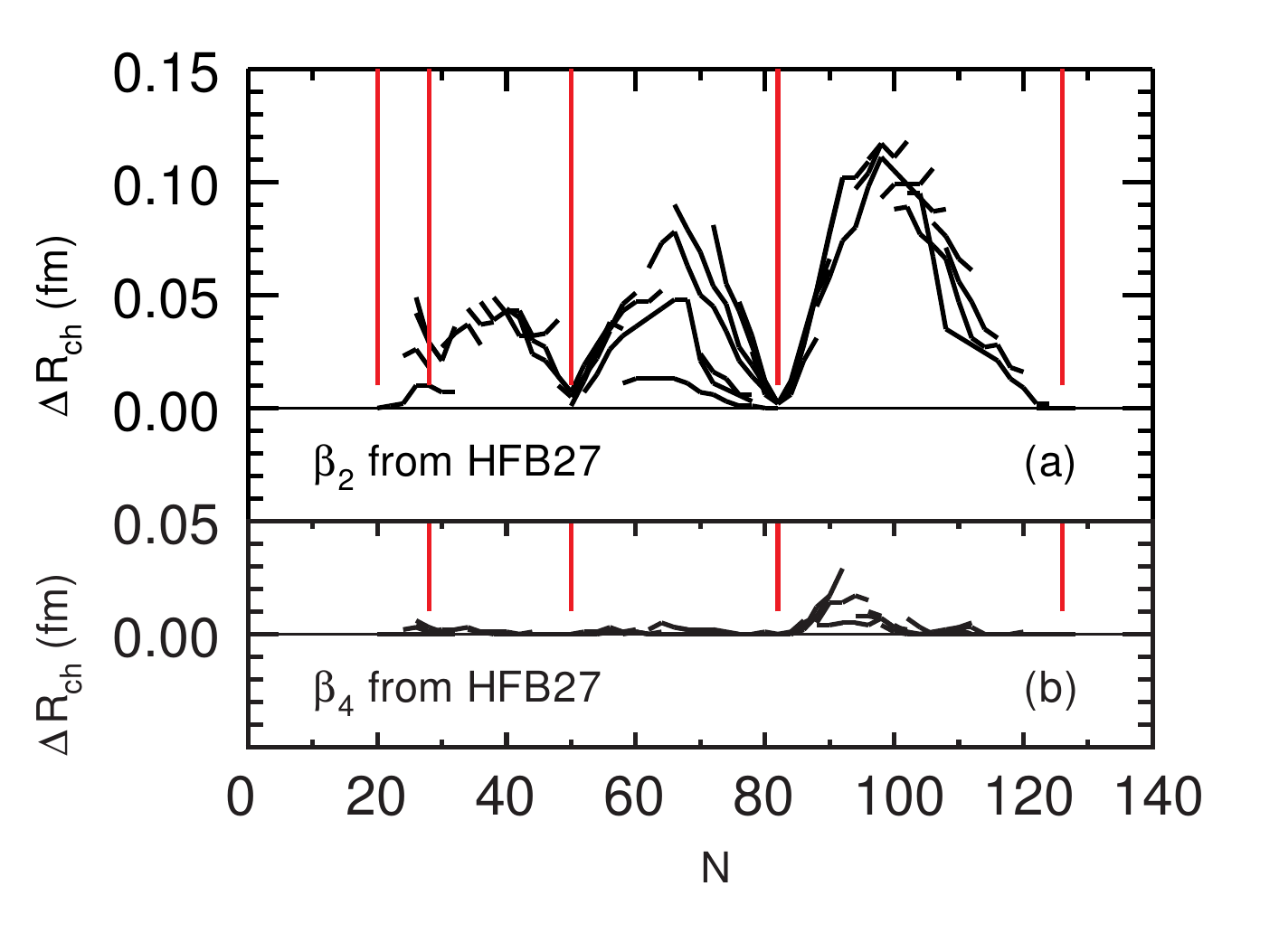}
\caption{ (a) Increase
in the rms charge radii calculated with Eq. (1) using
the $\beta_{2}$ values obtained in a deformed EDF calculation with
the HFB27 Skyrme functional. (b) Increase in the rms
charge radii calculated with Eq. (1) using the b$_{4}$
values obtained in a deformed EDF calculations with
the HFB27 Skyrme functional.
}
\label{ (5)] }
\end{figure}

\begin{figure}
\includegraphics[scale=0.5]{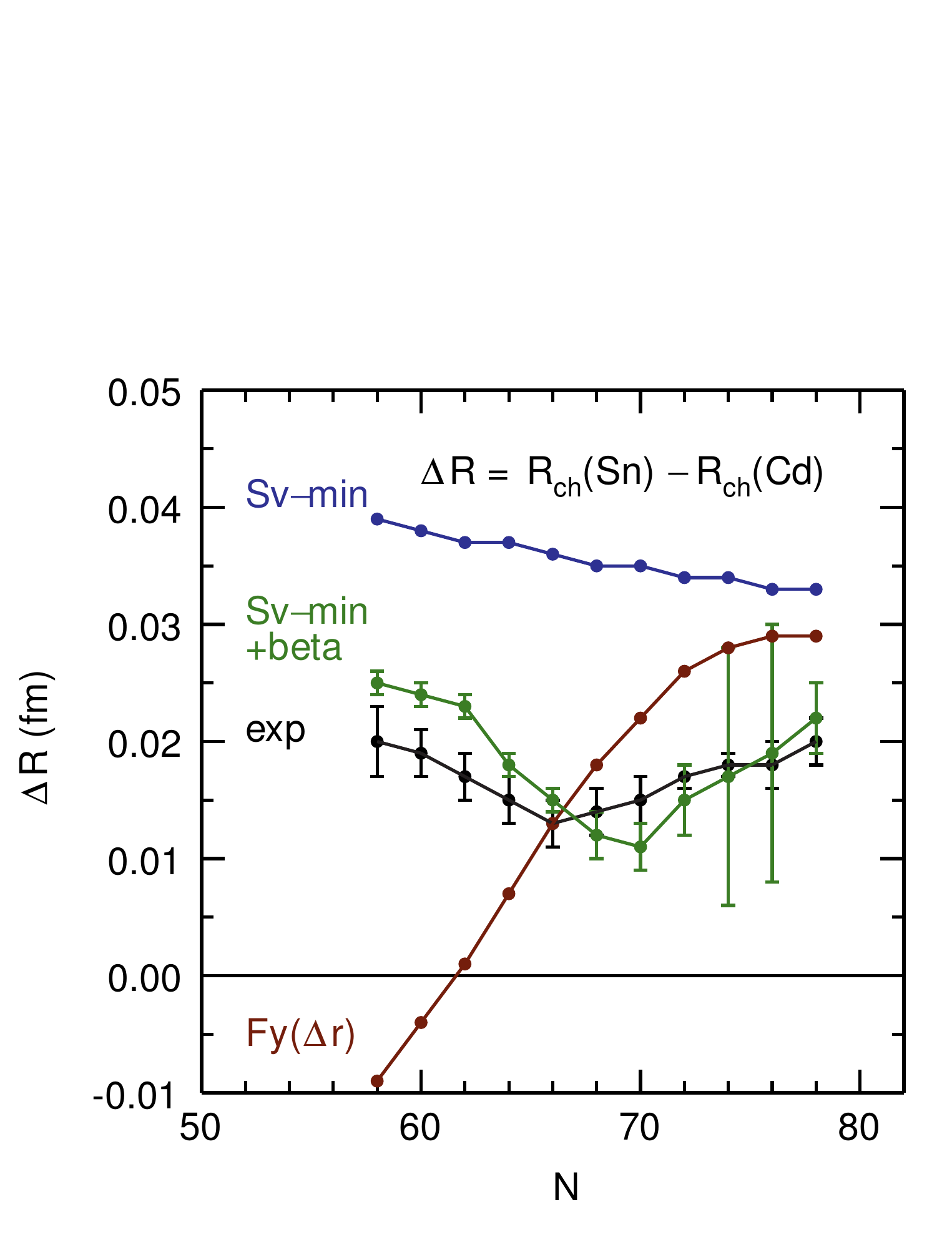}
\caption{Cd-Sn isotonic shifts. The experimental
results from \cite{cd18}, and \cite{sn19} are compared to the Sv-min and
Fr($\Delta$r) EDF calculations given in \cite{cd18} and \cite{sn19}.
}
\label{ (6)] }
\end{figure}

These large $  B(E2)  $ cannot be described by
calculations in the $  fp  $ model space \cite{lo21}.
They are a result of admixtures from configurations
with proton excitations
from the $  sd  $ shell to the $  pf  $ shell.
The configuration interaction (CI) calculations that include
these cross shell excitation are challenging.
In \cite{ca01} the ZBM2 Hamiltonian for the
$  (1s_{1/2},0d_{3/2},0f_{7/2},1p_{3/2})  $ model space
was used to calculate the rms charge radii of the  calcium isotopes
using harmonic-oscllator radial wavefunctions.
The number of protons excited from $  (1s_{1/2},0d_{3/2})  $ to
$  (0f_{7/2},1p_{3/2})  $ showed an odd-even effect. When these
orbital occupations were used with harmonic-oscllator radial
wavefunctions one obtained an increase in the rms charge radii with
odd-even oscillations that
were in qualitative agreement with experiment. However, in \cite{ro15}
when the orbital occupation numbers from these calculations
were used to constrain the spherical EDF calculations, the increase in the
rms charge radii was small compared to experiment.

To explore the $\beta_{2}^{2}$ contributions to the charge radii
we will use
the ZBM2-modified Hamiltonian for the $  (1s_{1/2},0d_{3/2},0f_{7/2},1p_{3/2})  $
model space as described in \cite{bi14}.
The corrections for the rms charge radii using
the $  B(E2)  $ values from these calculations
are shown by the blue line in panel (b) of Fig. (7).
We use effective charges of $  e_{p}=1.22  $ and $  e_{n}=0.78  $ from \cite{pi21}.
There is a disagreement with the calculated and experimental
$  B(E2)  $ for $^{46}$Ca. This can be traced to the location
of the $  2p-2h  $ proton intruder state that comes at
1.8 MeV in the calculations. Experimentally
it is observed at 2.4 MeV \cite{ca46}. ZBM2-modified Hamiltonian
was designed for the region of $^{40}$Ca. When it is used
for the region of $^{48}$Ca the proton shell gap is too small.
This can be fixed by adding a monopole term to the Hamiltonian
that moves the proton $  2p-2h  $ state in $^{46}$Ca up to 2.4 MeV.
The proton $  2p-2h  $ state in $^{48}$Ca is suggested to be at 4.28 MeV \cite{br98}
compared to  the calculated excitation energy of 4.45 MeV.
The results for the $\beta_{2}^{2}$ correction are shown by the
green line in panel (a) of Fig. (7).
EDF results
for $^{49,50}$Ca are discussed on the end of this paper.

In addition to the results for
even-even nuclei obtained with  $  B(E\lambda ,\uparrow)  $, we include
results for the odd-even nuclei obtained from
$$
<i\mid Q \cdot Q\mid i>\, = \frac{1}{(2J_{i}+1)} \displaystyle\sum _{c \ne i} <i||Q||c> <c||Q||i>,  
     \eqno({6})
$$
where $  Q  $ is the $  E2  $ operator.
For even-even nuclei this is $  B(E2,\uparrow)  $.
The sum for odd-even nuclei is typically over many excited states.
For example, in $^{43}$Ca, there are 11 excited states below 3.5 MeV that
contribute.

The odd-even effects in the orbital occupations and $  \beta _{2}^{2}  $
values are implicitly connected to the pairing interaction.
But the specific connections between the pairing oscillations
in the nuclear binding energies and the rms charge radii
depends upon details of the local nuclear structure.
There are many contributions to consider for rms radii:
the properties of the EDF functional,
orbital occupations, $\beta_{2}^{2}$ corrections, and possible
pairing-type additions to the EDF. Our results
suggest that the spherical EDF together with
the $\beta_{2}^{2}$ corrections are the most
important. But at a quantitative level of detail,
all must be considered.

\begin{figure}
\includegraphics[scale=0.7]{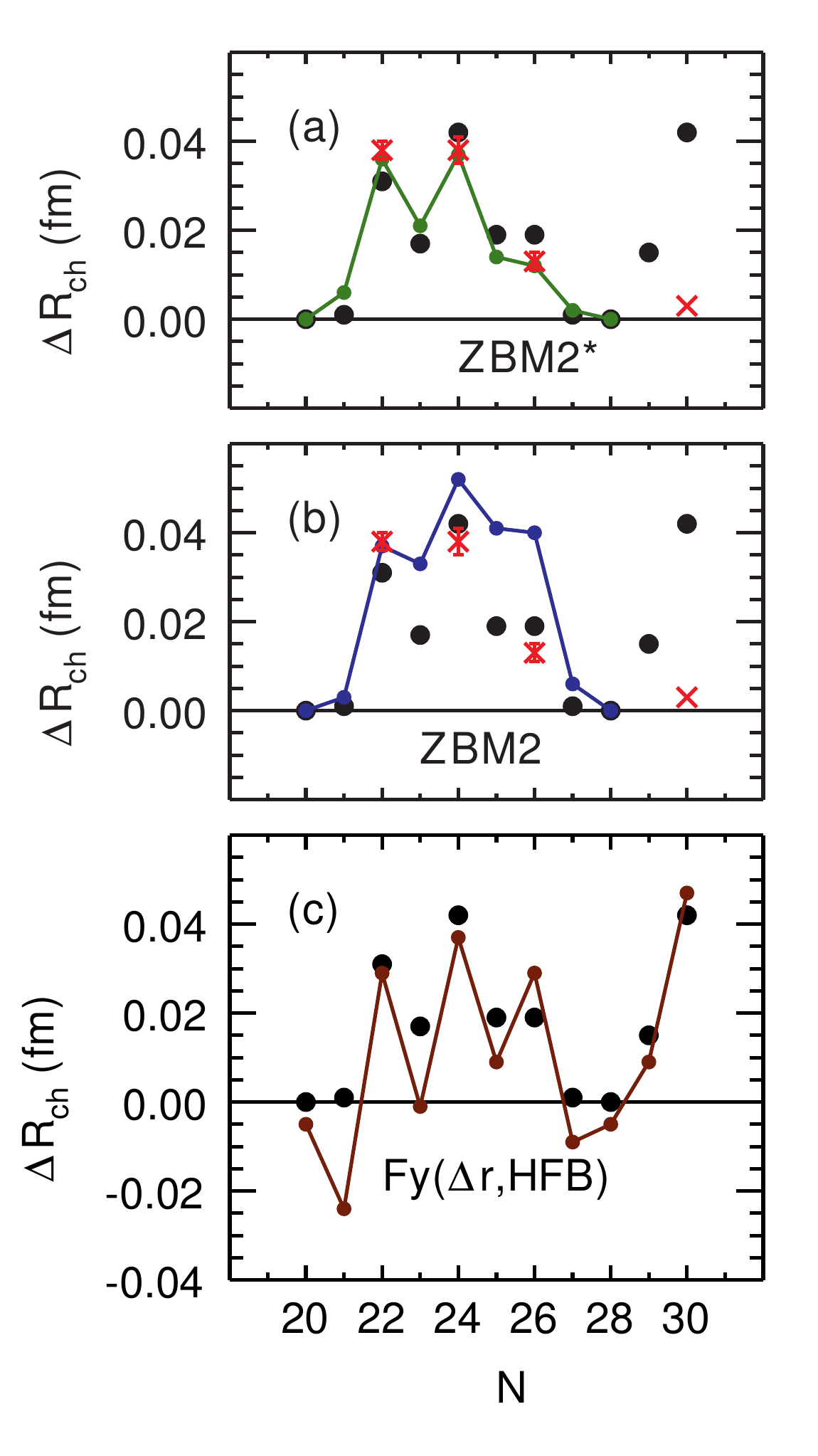}
\caption{$  \Delta R  _{ch}$ for the calcium isotopes.
The black circles are the
experimental results from \cite{ga16} with error bars about the
size of the circles.
The calculations shown in panel (c) are the
results from the Fy($\Delta$r,HFB) calculations given in \cite{mi19}.
The lines shown in panels (a) and (b) are
based on the $  B(E2)  $ calculations discussed in the
text.
The red crosses in panels (a) and (b) are based on experimental
$  B(E2)  $ values in the text.
}
\label{ (7) }
\end{figure}

\section{Isotonic Shifts for the $  N=28  $ Isotones}

The results for the $  N=28  $ isotones obtained with the GXPF1A Hamiltonian
for $  N=28  $ are shown by the green line in Fig. (8). We use effective charges
of $  e_{p}=1.22  $ and $  e_{n}=0.78  $ from \cite{pi21} and the
$\beta_{2}^{2}$ model of \cite{pi21} that includes the isovector term. One observes
odd-even oscillations in the calculated results.
The $\beta_{2}^{2}$ contribution only includes the 2$^{ + }_{1}$ state.
These are the calculations that were used to make the $\beta_{2}^{2}$
corrections for the connecting the mirror charge radii of
$^{54}$Ni and $^{54}$Fe to the slope parameter $  L  $ of
neutron equation of state in \cite{pi21}.
As noted in \cite{pi21}, for $^{54}$Fe there is additional
$  E2  $ strength at 2.8 MeV in the $  fp  $ calculation.
This is not included because it comes from coupling of
two-proton holes in $^{56}$Ni to the 2$^{ + }$ particle-hole excited
state in $^{56}$Ni at 2.7 MeV.

Charge radii calculated by the spherical EDF for $  N = 28  $
are linear as a function $  Z  $.
Thus, to obtain the $\beta^{2}$ correction from experiment we can
take the difference between the rms charge radius
for each isotope and that obtained from a linear extrapolation
between the experimental values for
$^{48}$Ca \cite{ga16} and $^{56}$Ni \cite{so22}. These results are shown by the black points
in Fig. (8).
More complete and
improved experimental results are needed for the odd-even isotopes.

\begin{figure}
\includegraphics[scale=0.6]{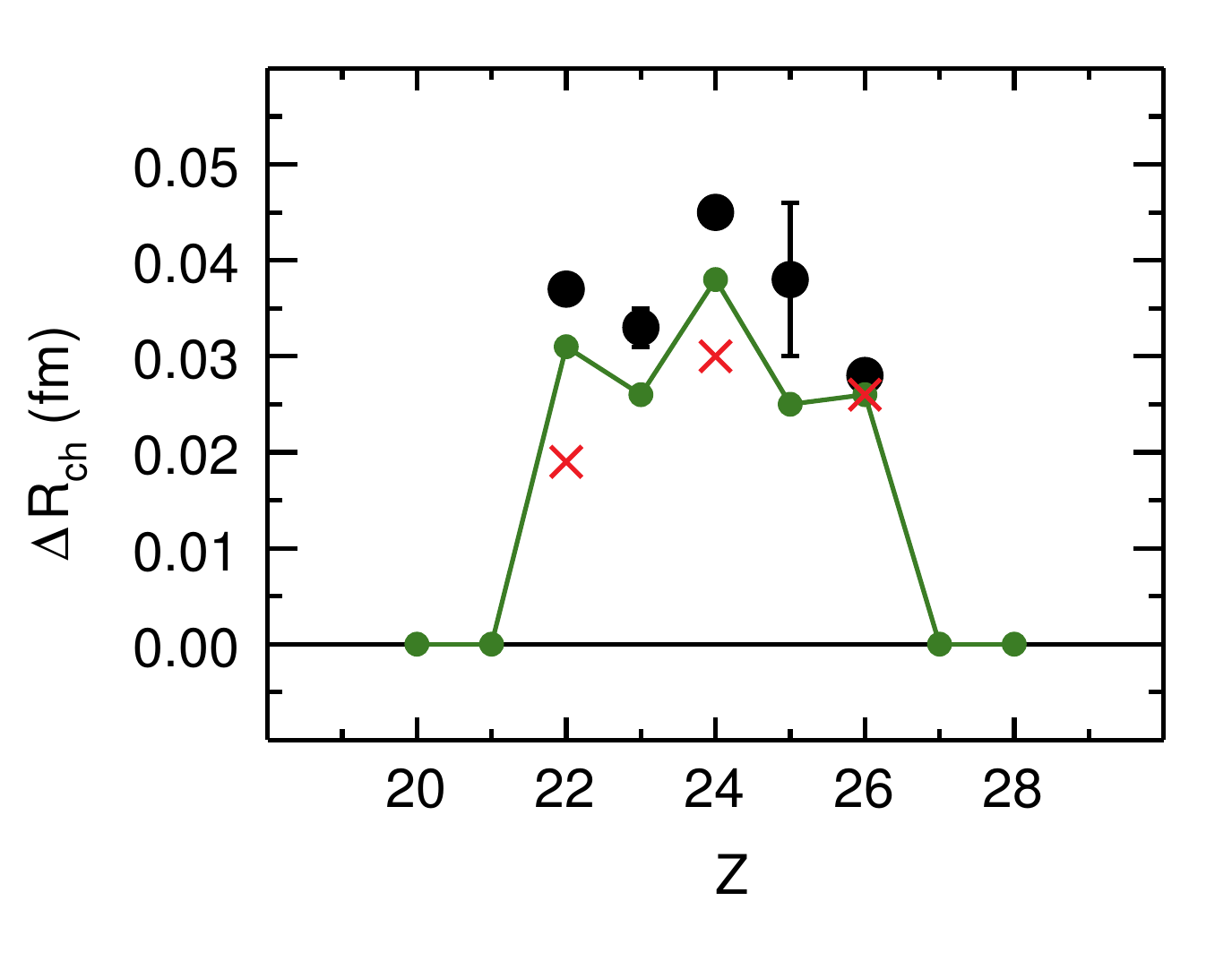}
\caption{$  \Delta R  _{ch}$ obtained from $\beta_{2}^{2}$ corrections for the
isotones with $  N=28  $ isotopes.
The red crosses are based on experimental $  B(E2,0^{+}_{1} \rightarrow 2^{+}_{1})  $
values.
The calculations shown by the green lines
are discussed in the text.
The experimental data differences between the
given isotope and a
a linear extrapolation between $^{48}$Ca \cite{ga16} and $^{56}$Ni \cite{so22}.
The data for Z = 22, 23, 24 and 26 are from  \cite{fr04}
using combined analyses of the muonic-X ray and electron scattering measurements,
where they are available.
Otherwise only the electron scattering data is used.
The charge radius for Z = 25 is from \cite{an13}.
Charge radii for Z = 21 and 27 are not experimentally known yet.
}
\label{ (8) }
\end{figure}

\begin{figure}
\includegraphics[scale=0.6]{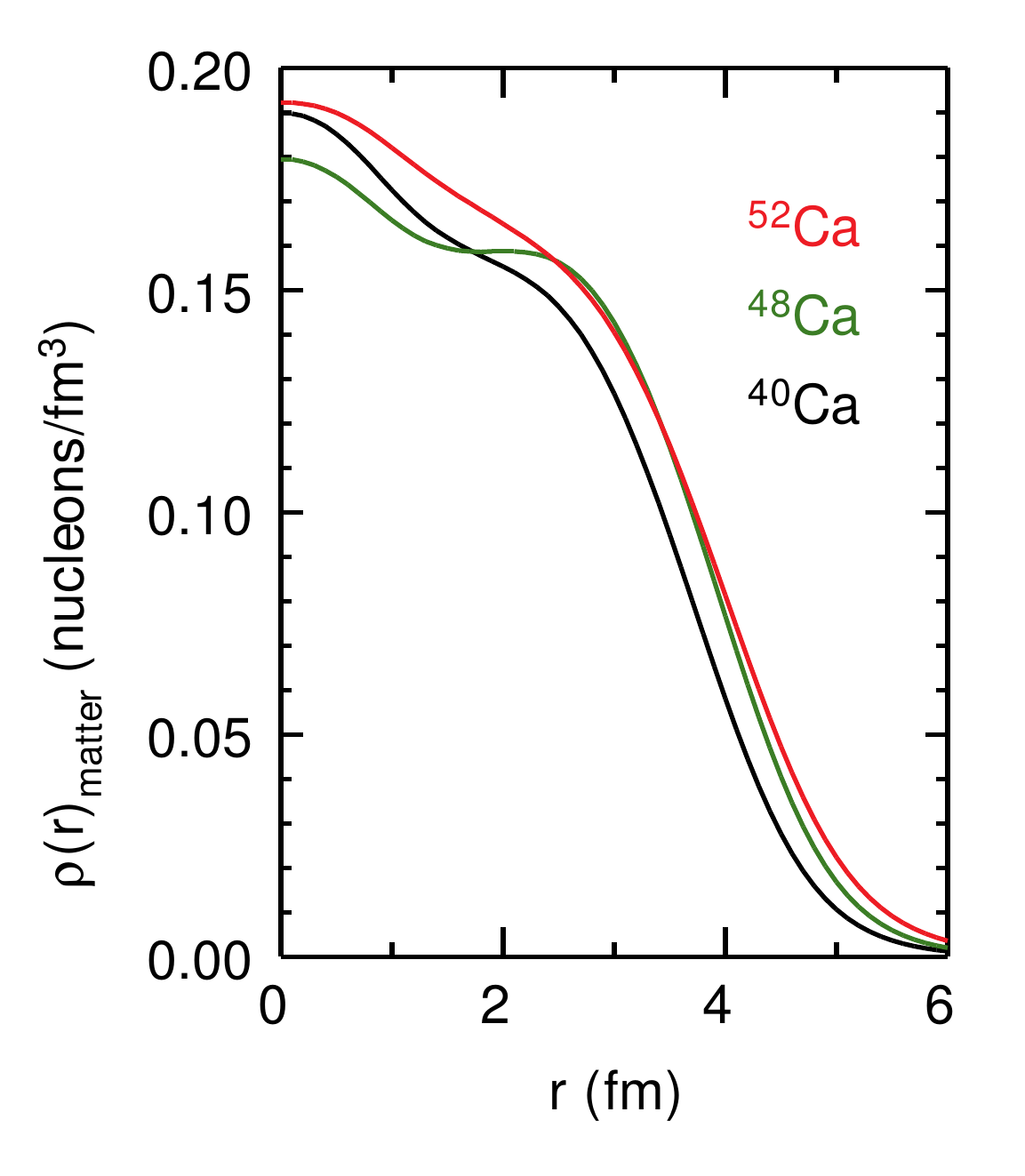}
\caption{Nuclear matter densities obtained with the $  s17  $ EDF
parameters.
}
\label{ (9) }
\end{figure}

The correlation between the difference in mirror charge radii and
the symmetry energy parameter $  L  $ is obtained with spherical
EDF \cite{br17} and CODF \cite{ya18} calculations
that are linear in between $^{48}$Ca and $^{56}$Ni. In addition, as discussed in
\cite{pi21}, one has to add
the $\beta_{2}^{2}$ corrections for $^{54}$Fe as shown in Fig. (8)
and for $^{54}$Ni.
The oscillations in the rms charge
radii are treated in
a different way in the Fayans-type functional by   adding
pairing-type functional with new
global parameters \cite{cd18}, \cite{sn19}, \cite{re17}. In the Fayans method some of the
correlation between the mirror charge radii and $  L  $
is lost \cite{re22}.

\section{Increase in Charge Radii After the Magic Numbers}

Spherical EDF with $\beta_{2}^{2}$ corrections do not account for the increase
in the rms charge radii just after the magic numbers shown in
the bottom of Figs. (2), (3) and (7). All of these
increases are associated with the sudden occupation
of the orbitals with an addition node given in Table I.
The experimental difference in the rms charge radii between $^{49}$Ca and $^{48}$Ca
is given in Table II and compared with results of some calculations.
The result for the covariant density functional (CODF) model
are taken as one quarter of the increase between $^{52}$Ca and
$^{48}$Ca shown in Fig. 24 of \cite{pe21}.
Nuclear matter densities for $^{40}$Ca, $^{48}$Ca and $^{52}$Ca
obtained with the $  s17  $ EDF parameter set are shown in Fig. (9).
The interior densities
show a large variation. In particular, the matter density
for $^{52}$Ca has a larger interior density due to the
node in the radial wavefunctions.
The "size" of the nucleus is determined by the interior saturation
conditions imposed by the EDF functional.
Perhaps the
present generation of EDF functionals are
not complex enough to correctly take into account
these variations in the interior shape.

It is noted in \cite{na84} that the orbitals given in Table I
are all associated
with a sudden change in the octupole instabilities via their
large $  B(E3)  $ values with higher orbitals that differ
by $  \Delta j=3  $, $  0g_{9/2}  $, $  0h_{11/2}  $, $  0i_{13/2}  $, and
$  0j_{15/2}  $, respectively (see Fig. 2 in \cite{na84}).
This results in a sudden increase
in the $  B(E3)  $  just after the magic numbers.
This may contribute to the increase in the $  \Delta R  _{ch}$ observed in
the bottom of
Figs. (2) and (3).
The $  E3  $ contribution from the $  (0p_{3/2})^{n}  $ configurations
for $^{48 + n}$Ca would increase linearly with $  n  $.
The $  E3  $ contribution from the $  (0d_{5/2})^{n}  $ configurations
for $^{90 + n}$Zr would be maximized in $^{96}$Zr where a peak
in the $  E3  $ contribution to $  \Delta R  _{ch}$ is observed in Fig. (4).
The octupole
contributions to the rms radii in an EDF model that includes
octupole degrees of freedom \cite{na84}, \cite{ca20} needs to be explored.

\begin{table}
\caption{Experimental increase in the rms charge radii after the magic numbers.}
\begin{center}
\begin{tabular}{|c|c|c|c|}
\hline
$  N  $ & orbital & Nuclei & $  \Delta R  _{ch}$ \\
\hline
      &         &        &   (fm)  \\
\hline
28 & 1p$_{3/2}$ & $^{49-48}$Ca & 0.015(3)  \\
50 & 1d$_{5/2}$ & $^{91-90}$Zr & 0.015(2)  \\
82 & 1f$_{7/2}$ & $^{133-132}$Sn &         \\
132 & 1g$_{9/2}$ & $^{209-208}$Pb & 0.009(2) \\
\hline
\end{tabular}
\end{center}
\end{table}

\begin{table}
\caption{Charge rms difference between $^{49}$Ca and $^{48}$Ca
in units of (fm)}
\begin{center}
\begin{tabular}{|l|c|c|}
\hline
 $  \Delta R  _{ch}$(exp) & Ref. & 0.015(3) \\
\hline
 Sv-min & \cite{mi19} & 0.008 \\
 $  s8  $ & \cite{br14}  &  0.009 \\
 $  s17  $ & \cite{br14} &  0.010 \\
 CODF & \cite{pe21} & 0.011 \\
\hline
 M3Y-P6a & \cite{na15} & 0.011 \\
 Fy($\Delta$r,BCS) & \cite{mi19} & 0.004 \\
 Fy($\Delta$r,HFB) & \cite{mi19} & 0.014 \\

\hline

\end{tabular}
\end{center}
\end{table}

\section{Conclusions}

In this paper we considered the $\beta_{ \lambda }^{2}$ corrections to
rms radii provided by the Bohr model.
We showed results for $\lambda$=2, 3 and 4.
The most important $\beta_{2}^{2}$ contribution
can be obtained from  experimental
$  B(E2)  $ value. When these $\beta_{2}^{2}$ corrections
are added to spherical EDF calculations,
the rms charge radii for even-even nuclei are in
reasonable agreement with experiment.
The major deviations between theory and experiment
appear just after the magic numbers.
We also used the experimental $  B(E2)  $ to
calculate the isotonic shift between the cadmium and tin
isotopes, the isotopic shifts of the calcium isotopes
and the isotonic shifts of the $  N=28  $ isotones.
The results are all in reasonable agreement with the data.

For the calcium isotopes and the $  N=28  $ isotones,
we used configuration-mixing calculations to obtain
the $  <Q \cdot Q>  $ corrections for both even-even and odd-even
nuclei. The calculated $  B(E2)  $ for even-even nuclei
are in reasonable agreement with experiment. The
evaluation of $  <Q \cdot Q>  $ for the odd-even nuclei
leads to odd-even oscillations in the rms charge
radii that are also in reasonably good agreement
with the data.

For the $\beta_{2}^{2}$ correction model, the primary remaining deviation between
experimental and theoretical rms charge radii
is for the rapid increase just after the
magic numbers. We suggest that this may
come form octupole deformations and/or
a limitation in the standard form
of the Skyrme-type EDF functionals.

This work was supported by NSF grant PHY-2110365 and NSF PHY 2111185.
We thank H. Nakada and W. Nazarwicz for discussions and suggestions.

\end{document}